# A strongly inhomogeneous superfluid in an iron-based superconductor

*D. Cho\*,[1], K.M. Bastiaans\*,[1], D. Chatzopoulos\*,[1], G.D. Gu[2], M.P. Allan[1]¶*

[1] Leiden Institute of Physics, Leiden University, Niels Bohrweg 2, 2333 CA Leiden, The Netherlands
[2] Condensed Matter Physics and Materials Science Department, Brookhaven National Laboratory, Upton, NY 11973, USA

\* These authors contributed equally to this work.    ¶ allan@physics.leidenuniv.nl

**Among the mysteries surrounding unconventional, strongly correlated superconductors is the possibility of spatial variations in their superfluid density. We use atomic-resolution Josephson scanning tunneling microscopy to reveal a strongly inhomogeneous superfluid in the iron-based superconductor $FeTe_{0.55}Se_{0.45}$. By simultaneously measuring the topographic and electronic properties, we find that this inhomogeneity in the superfluid density is not caused by structural disorder or strong inter-pocket scattering, and does not correlate with variations in Cooper pair-breaking gap. Instead, we see a clear spatial correlation between superfluid density and quasiparticle strength, putting the iron-based superconductors on equal footing with the cuprates and demonstrating that locally, the quasiparticles are sharpest when the superconductivity is strongest. When repeated at different temperatures, our technique could further help elucidate what local and global mechanisms limit the critical temperature in unconventional superconductors.**

Superconductivity emerges when electrons pair up to form so-called Cooper pairs and then establish phase coherence to condense into the macroscopic quantum state that is the superfluid. Cooper pairing is governed by the binding energy of the pairs, $\Delta_{PB}$, while the phase coherence or stiffness, governs the superfluid density, $n_{sf}$.[1,2] For conventional superconductors like aluminum or lead, the superfluid density is spatially homogeneous because the lattice constant is much smaller than the Cooper pair size of a few tens of nanometer, and because the large superfluid density guarantees a high phase stiffness.  In unconventional, strongly correlated superconductors the situation is very different for the following reasons: (i) the Cooper pair size, roughly given by the coherence length, is smaller; (ii) the superfluid density is smaller (iii), more disorder exists due to dopant atoms or intrinsic tendencies for phase separation or charge order; and (iv) the superconducting gap changes sign. Despite much progress[3,4], we lack a theoretical understanding of these strongly correlated superconductors. It has long been proposed that there can, in principle, exist spatial variations of the superfluid density.[5-8] Very similar ideas have been discussed thoroughly in the context of superconductor-insulator transitions[8-11], or Bose-Einstein condensation of electronic liquids.[12] However, little is known about the local physics in such systems because of the technical challenges associated with visualizing the superfluid density on the atomic scale, especially when simultaneously probing the density of states to investigate the origin of the inhomogeneity.

The pair-breaking gap and the superfluid density should be accessible through two distinct spectroscopic signatures in a tunneling contact between superconductors (Fig 1a). The first one is visible in the single-particle channel, where Bogoliubov quasiparticles with energies larger than the pair-breaking gaps transport the charge, as shown in Fig 1b. In the case of the scanning tunneling microscopy (STM) configuration relevant to this Letter, one of the superconductors is the tip with gap $\Delta_{PB,t}$ and the other is the sample with gap $\Delta_{PB,s}$; leading to a total gap of energy $2(\Delta_{PB,s} + \Delta_{PB,t})$ (Fig 1c). The second spectroscopic feature is at bias energies close to the Fermi energy, where one can access the Cooper-pair channel which yield information about the superfluid density. Voltage-biased Josephson tunneling in our STM configuration differs somewhat from the case of planar junctions: the capacitive energy $E_C$ is much bigger than the Josephson energy, $E_J$, turning the environmental impedance into a relevant quantity, and, in our case, the thermal energy is relatively high. Figure 1d shows the equivalent circuit for a generic junction in an STM environment.

We calculate the current-voltage characteristics of Josephson tunneling based on two different theoretical frameworks: IZ and P(E). The former, named after its developers Ivanchenko and Zil'berman, models the environment as Ohmic and assumes that the thermal energy exceeds the Josephson energy.[13] The latter, named after the probability function central to the theory, is a quantum mechanical treatment of Cooper pair tunneling in ultra-small junctions.[14,15] For our specific configuration, the qualitative predictions from both theoretical descriptions are similar: a Josephson current flows at small bias, with a maximum within a few microvolts around the Fermi energy (Fig 1e), reflected in a conductance spectrum that shows a peak at zero applied bias. The maximum Josephson current (arrow in Fig 1e) is proportional to the square of the critical current $I_C$ of the junction (see Supplementary Information). The superfluid density $n_{sf}$ can then be extracted by using $n_{sf} \propto (I_C R_N)^2$, where $R_N$ is the normal state resistance. Spatially imaging the superfluid density using such techniques[16] has thus far only been achieved on a cuprate sample with a resolution of ~1 nm, by exfoliating pieces of the sample onto the tip[17], leading to the discovery of a pair density wave, and with atomic resolution on Pb(111), using the sample material to coat the tip.[18]

In the present study, we investigate the unconventional iron-based superconductor FeTe$_{0.55}$Se$_{0.45}$. Iron-based superconductors are moderately to strongly correlated, with Hund's rule and orbital selectivity playing important roles.[19,20] We chose FeTe$_{0.55}$Se$_{0.45}$ because it encompasses the key properties of unconventional superconductivity and because its nodeless gap structure[21,22] and the possibility to scan at low junction resistances facilitate the Josephson experiments described below. FeTe$_{0.55}$Se$_{0.45}$ is considered not to be in the dirty BCS limit and has a low average superfluid density similar to cuprate high temperature superconductors.[23,24] We cleave the single crystals at 30 K and insert the samples into our cryogenic STM with rigorous electronic filtering. The topograph (Fig. 2a) shows atomic resolution and contrast differences that stem from the tellurium or selenium inhomogeneities. We use a mechanically sharpened platinum iridium wire with its apex coated with lead, which is a s-wave superconductor with a relatively large gap of ~1.3 meV .[18,25] We characterize and test its properties on a atomically flat Pb(111) surface (see Supplementary Information).

These preparations enable us to acquire Josephson tunneling spectra and maps on FeTe$_{0.55}$Se$_{0.45}$. Figure 2 shows current and differential conductance spectra acquired at the location marked by a cross in Fig. 2a. The data agrees well with expectations from the IZ and P(E) models. Note that it reproduces small oscillation features seen previously on elemental superconductors and explained by a tip-induced antenna mode.[18,26] We further note a small kink in the Josephson current (arrow in Fig 2b) that might be related to the coupling of a bosonic mode.

In Figs 3a and 3b, we show an atomic-resolution Josephson map, extracted from ~16'000 individual spectra, and the topographic image, registered to each other on the atomic scale (see Supplementary Information). The most striking finding of our experiment is the large change of the superfluid density over length scales of the order of the coherence length, a few nanometers. We show in Fig. 3c a series of individual raw spectra normalized by the normal state resistance to illustrate these changes. The inhomogeneities are not periodic; a possible pair density wave superimposed is below our sensitivity. Our setup allows us to measure topographic and electronic properties in the same field of view and thus investigate possible causes for the inhomogeneous superfluid. The most obvious possible causes might be structural disorder and strong quasiparticle scattering. The structural disorder stems from the effective FeSe and FeTe alloying that is clearly visible in the topographic images (Figs 2a, 3a). Surprisingly, the changes in the superfluid density are not correlated to these structural features, with the exception of a few impurity atoms that lead to a strong suppression. The strength of the quasiparticle scattering is visible in quasiparticle interference (QPI) pattern and is dominated by inter-pocket scattering in FeTe$_{0.55}$Se$_{0.45}$.[21] We identify areas of strong scattering with red contours in Fig. 3d. Again, there is no correlation to the superfluid density. We cannot exclude that the superfluid density is influenced by potential scatterers not visible in our measurement, remnant short range magnetic order, or possible phase separations at higher energies. However, the fact that such prominent effects as the structural disorder and well as the inter-pocket QPI do not influence the superfluid indicates that the inhomogeneity in the superfluid density is intrinsic.

We now return to the relation between the pair-breaking gap and the superfluid density. We extract the pair-breaking gap energy, as well as the height of the coherence peaks, which will prove to be important later, by fitting the coherence peak of each spectrum. Figure 3e shows the gap map for the same field of view as the Josephson map; the gap variations agree with previous reports.[27] It is clear that the pair-breaking gap is independent of the superfluid density. Instead we find a correlation to the quasiparticle character, as described below.

In unconventional superconductors, there is a recurring theme that connects quasiparticle excitation line-shapes with the presence of superconductivity: Photoemission demonstrated that the incoherent quasiparticles in the normal state become coherent below the critical temperature.[22,28] STM showed Bogoliubov QPI at low energies which are even sharper than theory would predict, but vanish well below the gap energy.[29,30] These measurements suggest a remarkable relation between the average quasiparticle excitation spectrum and superconductivity, but lack any notion of a possible inhomogeneous character in unconventional superconductors. While recently a relation between superfluid density and quasiparticle character has been conjectured to hold also locally for single-layer cuprates[31], direct experimental evidence is thus far missing. Our measurement allows us to extract the quasiparticle strength (QPS), which we define phenomenologically as the height of the coherence peak (Fig. 3f), and relate it directly to the superfluid density at the same location. Indeed, we find a striking correlation between the superfluid density and the QPS over the whole field of view, with a linear correlation coefficient of 0.58 (Fig. 4). While there is no theory yet that can explain such a phenomenology, our finding has implications both for iron-based superconductors and for unconventional superconductors in general. It demonstrates a similarity between FeTe$_{0.55}$Se$_{0.45}$ and the cuprates, where such an effect is present around magnetic impurities, and contrasts both of them with the conventional superconductors where it is absent.[18] Further, it points towards a local mechanism behind the relation found by photoemission, a condition fulfilled by pinned thermal phase fluctuations and glassy superconductivity.

In summary, we have detected and directly imaged a strongly inhomogeneous superfluid and simultaneously measured the electronic and topographic properties in the same field of view, with atomic resolution. We found that the superfluid inhomogeneity is not caused by the structural disorder resulting from the Se/Te alloying, by the inter-pocket scattering, or by the variations of the pair-breaking gap. Instead, the superfluid density shows strong positive correlation with the sharpness of the quasiparticle peak: Superconductivity is needed for coherent quasiparticles, locally on the length scale of cooper pairing. It will be instructive to investigate the superfluid density in other materials using our techniques, including superconductor-insulator transitions, disordered conventional superconductors, or twisted bilayer graphene.[32,33] Lastly, we anticipate that future temperature-dependent superfluid density and gap measurements[27] will elucidate what local and global mechanisms limit $T_c$ in unconventional superconductors.

*Acknowledgements*

We are grateful to J.C. Davis, S.D. Edkins, K.J. Franke, H. Grabert, M.H. Hamidian, K. Heeck, M. Leeuwenhoek, G. Verdoes and J. Zaanen for valuable discussions. This work was supported by the European Research Council (ERC StG SpinMelt) and by the Netherlands Organization for Scientific Research (NWO/OCW), as part of the Frontiers of Nanoscience program, as well as through a Vidi grant (680-47-536). GDG is supported by the Office of Basic Energy Sciences, Materials Sciences and Engineering Division, U.S. Department of Energy (DOE) under Contract No. de-sc0012704.

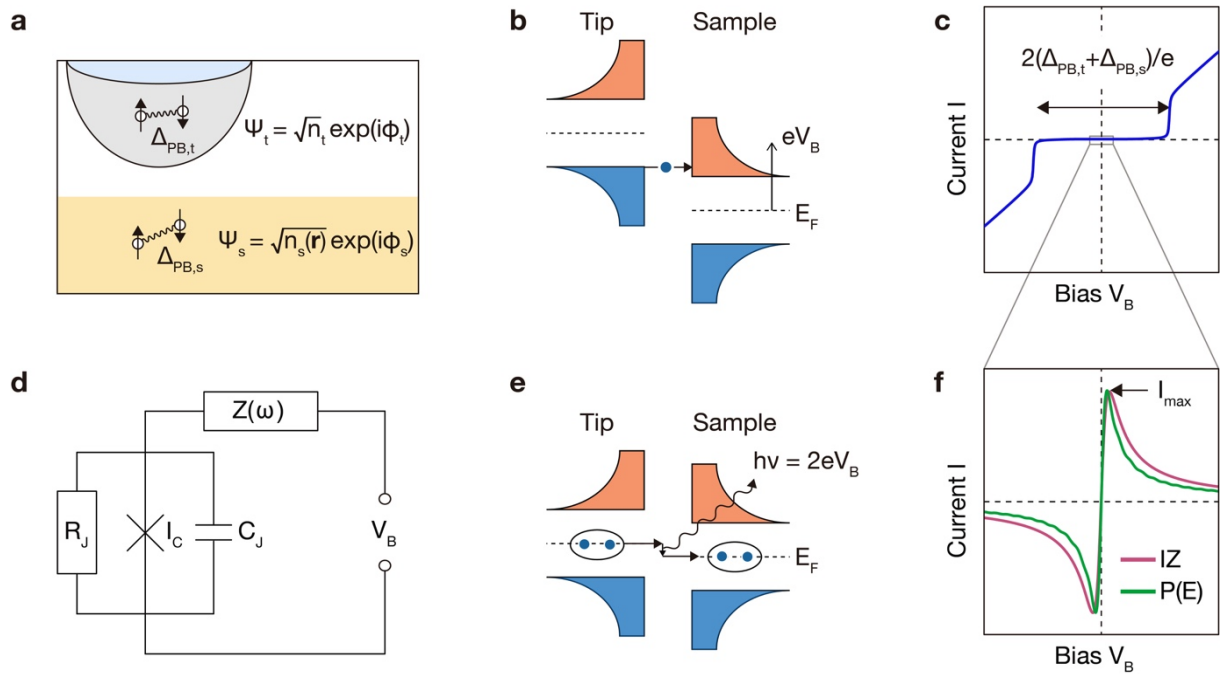

**Figure 1. Principles of Josephson Scanning Tunneling Microscopy. a.** Schematic of the Josephson junction consisting of tip (t) and sample (s). **b.** Schematic energy diagram of quasiparticle tunneling between tip and sample. Black lines indicate the density of states (horizontal axis) as a function of energy (vertical axis); filled/empty states are denoted with blue/red; dashed lines indicate the Fermi level $E_F$. When the voltage bias $V_B$ is larger than $(\Delta_{PB,s} + \Delta_{PB,t})/e$, quasiparticles can tunnel. **c.** Current-voltage I(V) characteristic curve for quasiparticle tunneling. **d.** Equivalent circuit diagram of the Josephson junction, $Z(\omega)$ represents the electromagnetic environment. **e.** Schematic of inelastic Cooper-pair tunneling in a Josephson junction. Cooper pairs interacts with the environment by emitting energy (wavy arrow) and subsequently tunnels across the junction. **f.** Simulated IV curves from Cooper-pair tunneling. Both curves exhibit a maximum $I_{max}$ at finite bias which is proportional to $I_C^2$.

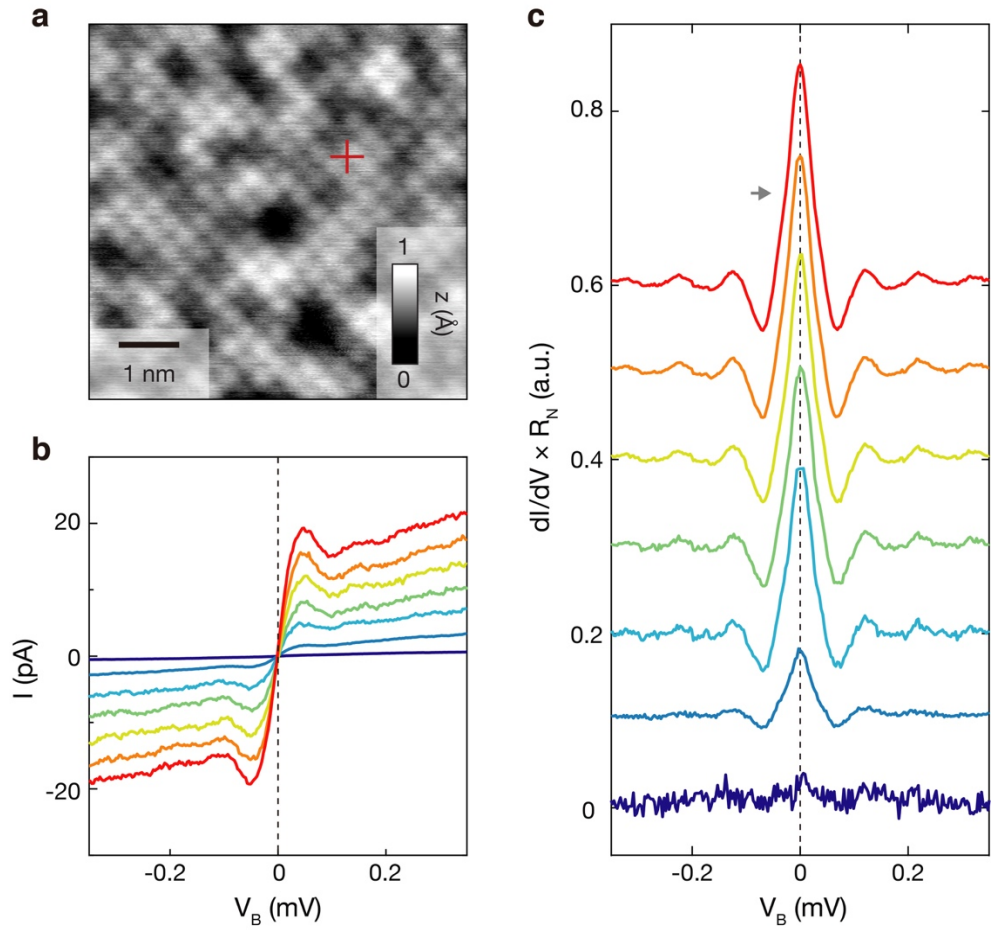

**Figure 2**. **Josephson tunneling spectra on FeTe$_{0.55}$Se$_{0.45}$. a.** Atomically resolved topographic image. Brighter (darker) atoms correspond to Te (Se). **b.** Current-voltage characteristic at the location indicated with a red cross in A, for different junction resistances from 14.0 (dark blue) to 0.4 (red) MOhm. **c**. Differential conductance spectra multiplied by the normal state resistance for the IV curves shown in b.

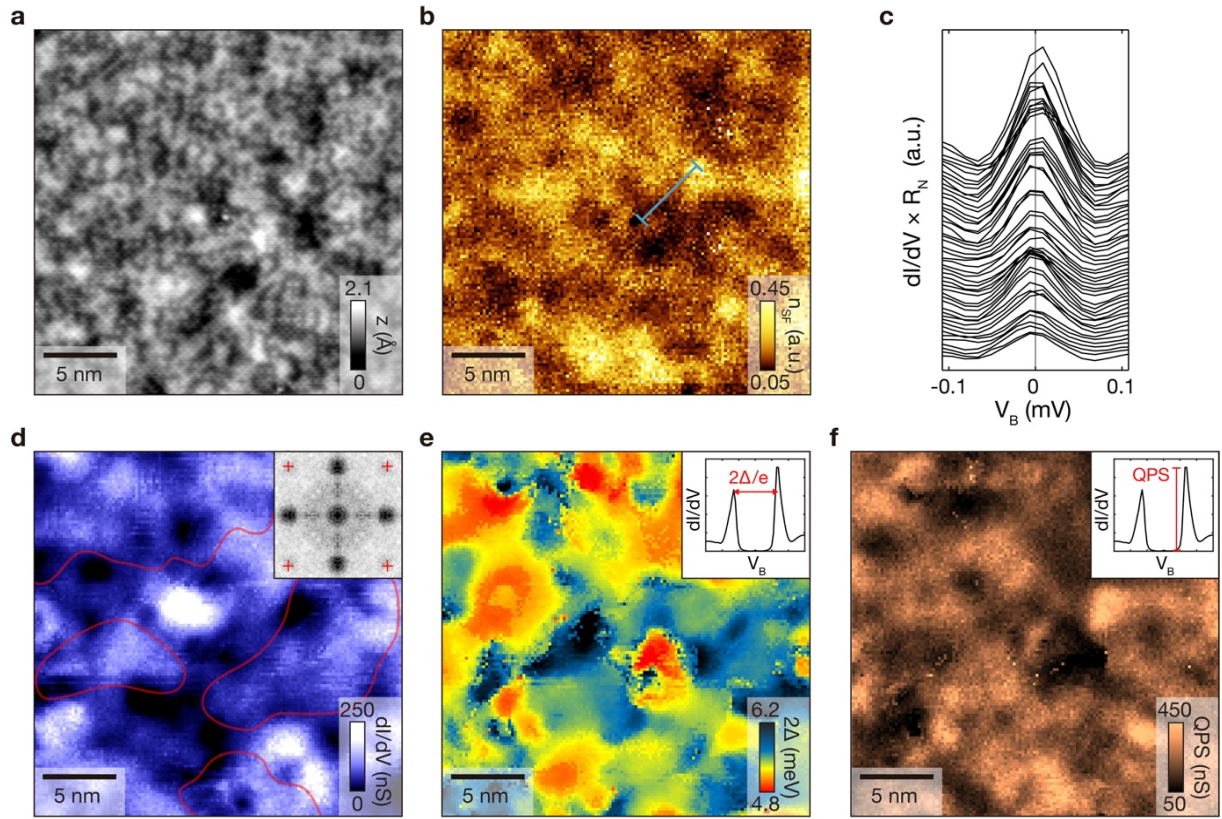

**Figure 3. Visualizing the superfluid density on FeTe$_{0.55}$Se$_{0.45}$.** **a**. 25x25 nm$^2$ topographic image of FeTe$_{0.55}$Se$_{0.45}$. **b.** Spatially resolved map of the superfluid density. **c.** Series of differential conductance spectra multiplied by the normal state resistance around E$_F$ along the blue line in b. **d**. Conductance map at $V_B$ =+3.6 mV. Areas with strong quasiparticle interference patterns are marked by red contours. Inset: Fourier transform, with crosses at the Bragg peak locations. **e.** Pair-breaking gap map. **f.** Coherence peak-height map (QPS), extracted simultaneously with the pair-breaking gap. All maps in b-f were obtained in the same field of view as the topography in a.

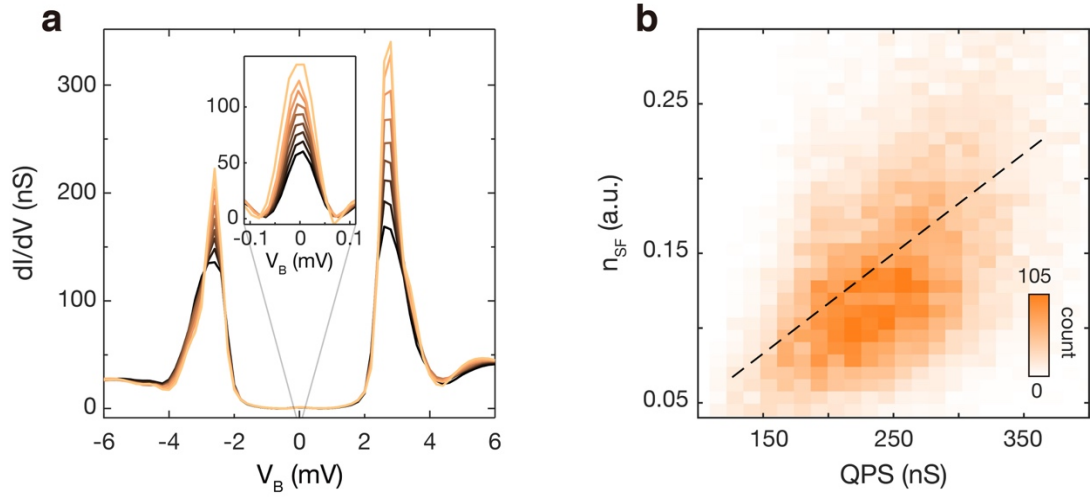

**Figure 4. Correlation between superfluid density and coherence peak-height.** **a.** Sorted spectra of the coherence peak-height and the zero-bias Josephson peak (inset, using different set-up conditions). Spectra were sorted by binning of the superfluid density map shown in Fig. 3b. The colors correspond to the quasiparticle strength in Fig. 3f. **b.** Correlation between coherence peak-height and superfluid density yielding a correlation factor of 0.58 (dashed line).